\documentclass[conference,letterpaper]{IEEEtran}

\IEEEoverridecommandlockouts

\usepackage{setspace}
\usepackage{cite}
\usepackage{amsmath,amssymb,amsfonts,mathtools,bm}
\usepackage{amsthm}
\usepackage{algorithm}
\usepackage{algpseudocode}
\usepackage{graphicx}
\usepackage{textcomp}
\usepackage{xcolor,float}
\usepackage{tabularx}
\usepackage{diagbox}
\usepackage{svg}
\usepackage{booktabs}
\usepackage[bookmarks=false,hidelinks]{hyperref}
\usepackage{makecell}
\usepackage{caption}
\usepackage{sidecap}
\usepackage{microtype}
\usepackage{multirow}
\usepackage{adjustbox}
\usepackage{colortbl}
\usepackage{url}
\usepackage{balance}
\usepackage{bbding}
\usepackage{array}
\usepackage{subcaption}
\usepackage{tikz}
\usepackage{changes}

\hypersetup{
    bookmarks=false,
    pdfpagemode=UseNone
}

\makeatletter
\let\ps@IEEEtitlepagestyle\ps@empty
\makeatother
\definecolor{lime}{HTML}{A6CE39}
\DeclareRobustCommand{\orcidicon}{%
    \begin{tikzpicture}
    \draw[lime, fill=lime] (0,0)
    circle [radius=0.16]
    node[white] {{\fontfamily{qag}\selectfont \tiny ID}};
    \draw[white, fill=white] (-0.0625,0.095)
    circle [radius=0.007];
    \end{tikzpicture}
    \hspace{-2mm}}

\foreach \x in {A, ..., Z}{%
    \expandafter\xdef\csname orcid\x\endcsname{\noexpand\href{https://orcid.org/\csname orcidauthor\x\endcsname}{\noexpand\orcidicon}}
}

\allowdisplaybreaks
\renewcommand{\baselinestretch}{1}

\begin{document}

\title{RadioVIL: Anomaly-Aware Diffusion Models for Radio Map Inpainting and Zero-Shot Vehicle Localization}

\author{
\IEEEauthorblockN{
Ruixin Zhao\IEEEauthorrefmark{1},
Xiucheng Wang\IEEEauthorrefmark{2},
Qiming Zhang\IEEEauthorrefmark{3},
Nan Cheng\IEEEauthorrefmark{2},
Ruijin Sun\IEEEauthorrefmark{2}, and
Conghao Zhou\IEEEauthorrefmark{2}
}

\IEEEauthorblockA{
\IEEEauthorrefmark{1}School of Computer Science and Technology, Xidian University, Xi'an, 710071, China\\
\IEEEauthorrefmark{2}State Key Laboratory of ISN and School of Telecommunications Engineering, Xidian University, Xi'an, 710071, China\\
\IEEEauthorrefmark{3}School of Artificial Intelligence, Xidian University, Xi'an, 710071, China\\
Email: \{24179100115, xcwang\_1, 23009200991\}@stu.xidian.edu.cn, dr.nan.cheng@ieee.org,\\
sunruijin@xidian.edu.cn, conghao.zhou@ieee.org
}
}

\maketitle
\thispagestyle{empty}

\begin{abstract}
High-precision radio map construction is essential for emerging 6G Integrated Sensing and Communication (ISAC) applications, including digital twins and intelligent transportation. However, existing deep learning methods predominantly treat this as a pure image completion task, resulting in over-smoothed reconstructions that fundamentally erase high-frequency scattering signatures of dynamic physical entities such as hidden vehicles. To overcome this, we propose RadioVIL, an efficient two-stage framework that reformulates joint radio map inpainting and zero-shot vehicle localization as a prior-guided physical inverse problem. Specifically, we first train a Denoising Diffusion Probabilistic Model (DDPM) to capture the structural generative prior of the environment. During inference from highly sparse measurements, we employ a Diffusion-based Mediating Intermediate Layer Optimization (DMILO) algorithm. By optimizing an $L_1$-regularized sparse deviation term, DMILO mathematically isolates vehicle scattering anomalies layer-by-layer without unfolding the entire denoising chain. Extensive experiments demonstrate that while conventional reconstruction baselines fail to detect hidden vehicles, and the zero-shot diffusion baseline achieves only limited detection ability due to forced semantic harmonization, RadioVIL preserves authentic physical textures, yielding the best LPIPS of 0.0587 in our evaluation. Uniquely, it unlocks accurate zero-shot vehicle localization directly from sparse radio maps, securing a 75.20\% Recall and a 3.31-meter average error, paving a robust way for ISAC at the 6G edge.
\end{abstract}


\section{Introduction}
\label{sec:introduction}

The evolution toward 6G Integrated Sensing and Communication (ISAC) requires high-fidelity, environment-aware electromagnetic representations. Radio maps (RMs) link environmental geometry and materials to spatial signal strength, serving as digital twins for applications such as autonomous driving, zero-measurement path planning, and dynamic vehicle localization. However, crowdsourced sensing is limited by deployment costs, physical occlusions, and restricted access to drivable road regions. The resulting sparse measurements make signal-field reconstruction and dynamic target detection highly ill-posed.

Existing methods, from regression-based RadioUNet \cite{levie2021radiounet} to adversarial RME-GAN \cite{zhang2023rme}, mainly treat RM construction as pixel-level completion. Pixel-wise objectives such as Mean Squared Error favor overly smooth reconstructions. Although effective for macroscopic path-loss trends, they erase high-frequency local variations and vehicle scattering footprints, hindering hidden-target sensing.

Recently, Denoising Diffusion Probabilistic Models (DDPMs) \cite{ho2020denoising} have shown strong ability in capturing complex data manifolds. In wireless domains, diffusion architectures have been extended to dynamic map construction \cite{wang2024radiodiff}, Helmholtz equation-informed physical modeling \cite{wang2025radiodiffk2}, 3D environment synthesis \cite{wang2025radiodiff3d}, and accelerated inference trajectories \cite{wang2025radiodiffflux}. Efficient architectures such as RadioMamba have also been introduced to address the computational burden of high-resolution feature extraction \cite{jia2025radiomamba}.

Despite these advances, solving the joint inverse problem under extreme sparsity remains challenging. For instance, zero-shot inpainting methods such as RePaint \cite{Lugmayr2022RePaint} achieve strong visual completion but fail in physical RM reconstruction. RePaint enforces semantic harmonization and blends masked regions with the background. Since hidden vehicles act as physical scatterers that contradict the smooth structural prior, RePaint suppresses these anomalies and effectively ``washes out'' vehicle signatures. Moreover, enforcing physical constraints across the entire denoising graph incurs high memory cost and may disrupt the generative manifold.

To address this bottleneck, we propose RadioVIL, an efficient two-stage framework that reformulates RM inpainting and zero-shot vehicle localization as an anomaly-aware, prior-guided physical inverse problem. Instead of end-to-end hallucination or rigid pixel replacement, RadioVIL dynamically injects hard physical constraints into the generative process. A pre-trained DDPM provides a smooth structural prior, while a layer-wise optimization strategy \cite{dmilo} isolates scattering anomalies without unfolding the full denoising chain. By introducing an $L_1$-regularized sparse deviation term, RadioVIL absorbs high-frequency measurement residuals that contradict the generative prior, preserving authentic electromagnetic signatures of hidden targets. The main contributions are summarized as follows:

\begin{enumerate}
    \item We introduce a two-stage framework integrating DDPMs \cite{ho2020denoising} with Diffusion-based Mediating Intermediate Layer Optimization (DMILO) \cite{dmilo} to solve the severely ill-posed radio map inpainting and sensing problem.
    \item We propose an anomaly-aware layer-wise optimization strategy that injects hard physical constraints. By using an $L_1$-regularized sparse deviation term, RadioVIL isolates high-frequency scattering anomalies and overcomes the semantic over-smoothing flaw of RePaint and traditional deep learning methods.
    \item We design a zero-shot vehicle extraction mechanism based on spatial morphology and Connected Component Analysis (CCA), enabling hidden vehicle localization directly from isolated anomalies without paired labels or auxiliary detection networks.
\end{enumerate}

\section{Preliminaries and System Model}
\label{sec:preliminaries_system_model}

\subsection{Denoising Diffusion Probabilistic Models}

Denoising Diffusion Probabilistic Models (DDPMs) \cite{ho2020denoising} have recently emerged as state-of-the-art generative frameworks, demonstrating exceptional capabilities in data synthesis and wireless environment reconstruction \cite{wang2024radiodiff, wang2025radiodiff}. Given a complete ground-truth data distribution $\mathcal{D}$, a DDPM progressively corrupts a clean sample $\bm{x}_0 \sim \mathcal{D}$ into a sequence of noisy latent variables $\{\bm{x}_t\}_{t=1}^T$ through a predefined variance schedule $\{\beta_t\}_{t=1}^T$. By defining $\alpha_t = 1 - \beta_t$ and $\bar{\alpha}_t = \prod_{s=1}^t \alpha_s$, the forward marginal distribution admits a closed-form Gaussian transition:
\begin{equation}
q(\bm{x}_t | \bm{x}_0) = \mathcal{N}\left(\bm{x}_t; \sqrt{\bar{\alpha}_t}\bm{x}_0, (1-\bar{\alpha}_t)\mathbf{I}\right).
\label{eq:forward_marginal}
\end{equation}

To invert this forward diffusion corruption, a neural network $\bm{\epsilon}_\theta(\bm{x}_t, t)$, typically parameterized as a U-Net architecture, is optimized to predict the injected Gaussian noise at each timestep. The network is trained by minimizing the variational lower bound (VLB) objective:
\begin{equation}
\mathcal{L}_{prior}(\theta) = \mathbb{E}_{\bm{x}_0, \bm{\epsilon}, t} \left[ \left\| \bm{\epsilon} - \bm{\epsilon}_\theta(\sqrt{\bar{\alpha}_t}\bm{x}_0 + \sqrt{1-\bar{\alpha}_t}\bm{\epsilon}, t) \right\|_2^2 \right].
\label{eq:training_loss}
\end{equation}
Once converged offline, the frozen denoising network effectively encapsulates the structural prior of the physical environment, serving as a powerful generative dictionary that represents the physically plausible data manifold.

\subsection{System Model and Problem Formulation}

Let the physical space be discretized into a two-dimensional grid of size $H \times W$. We define the underlying true radio map, which contains complete environmental information including dynamic hidden vehicles, as $\mathbf{R}^* \in \mathbb{R}^{H \times W}$. The physical environment is characterized by strict topological constraints. To formally represent the sensing capability, we define a binary physical mask matrix $\mathbf{M} \in \{0, 1\}^{H \times W}$. According to realistic urban configurations, the mask is constructed piecewise:
\begin{equation}
\mathbf{M}(p) =
\begin{cases}
1, & p \in \Omega_{bldg} \cup \Omega_{bs} \cup \Omega_{edge} \cup \Omega_{rand} \\
0, & p \in \Omega_{road} \cup \Omega_{veh}
\end{cases},
\label{eq:mask_definition}
\end{equation}
where observable regions include static buildings $\Omega_{bldg}$, base stations $\Omega_{bs}$, narrow road-edge observation bands $\Omega_{edge}$, and 5\%--10\% random free-space samples $\Omega_{rand}$. Drivable roads and potential vehicle regions are masked out to establish a zero-knowledge zone for vehicle locations.

To account for hidden vehicles, we decompose the complete radio map into a smooth background and sparse vehicle-induced scattering footprints:
\begin{equation}
\mathbf{R}^* = \mathbf{R}_{bg}^* + \mathcal{K}(\bm{\nu}_{veh}^*),
\label{eq:radiomap_decomposition}
\end{equation}
where $\mathbf{R}_{bg}^*$ denotes the static-environment background, $\bm{\nu}_{veh}^*$ represents sparse vehicle-induced scattering sources, and $\mathcal{K}(\cdot)$ is an effective scattering footprint operator. The sparse observation is then modeled as
\begin{equation}
\bm{y} = \mathbf{M} \odot \left( \mathbf{R}_{bg}^* + \mathcal{K}(\bm{\nu}_{veh}^*) \right) + \bm{\epsilon},
\label{eq:mask_hadamard}
\end{equation}
where $\bm{\epsilon}\sim\mathcal{N}(0,\sigma_n^2)$. Although vehicle pixels are masked, their scattering footprints still perturb observable road-edge, base-station, and random samples. Hidden vehicles are therefore inferred from residuals on $\{p|\mathbf{M}(p)=1\}$ rather than direct observations, preserving the strict zero-knowledge assumption while retaining indirect physical observability.

\begin{figure}[!t]
    \centering
    \begin{subfigure}[b]{0.48\columnwidth}
        \centering
        \includegraphics[width=\textwidth]{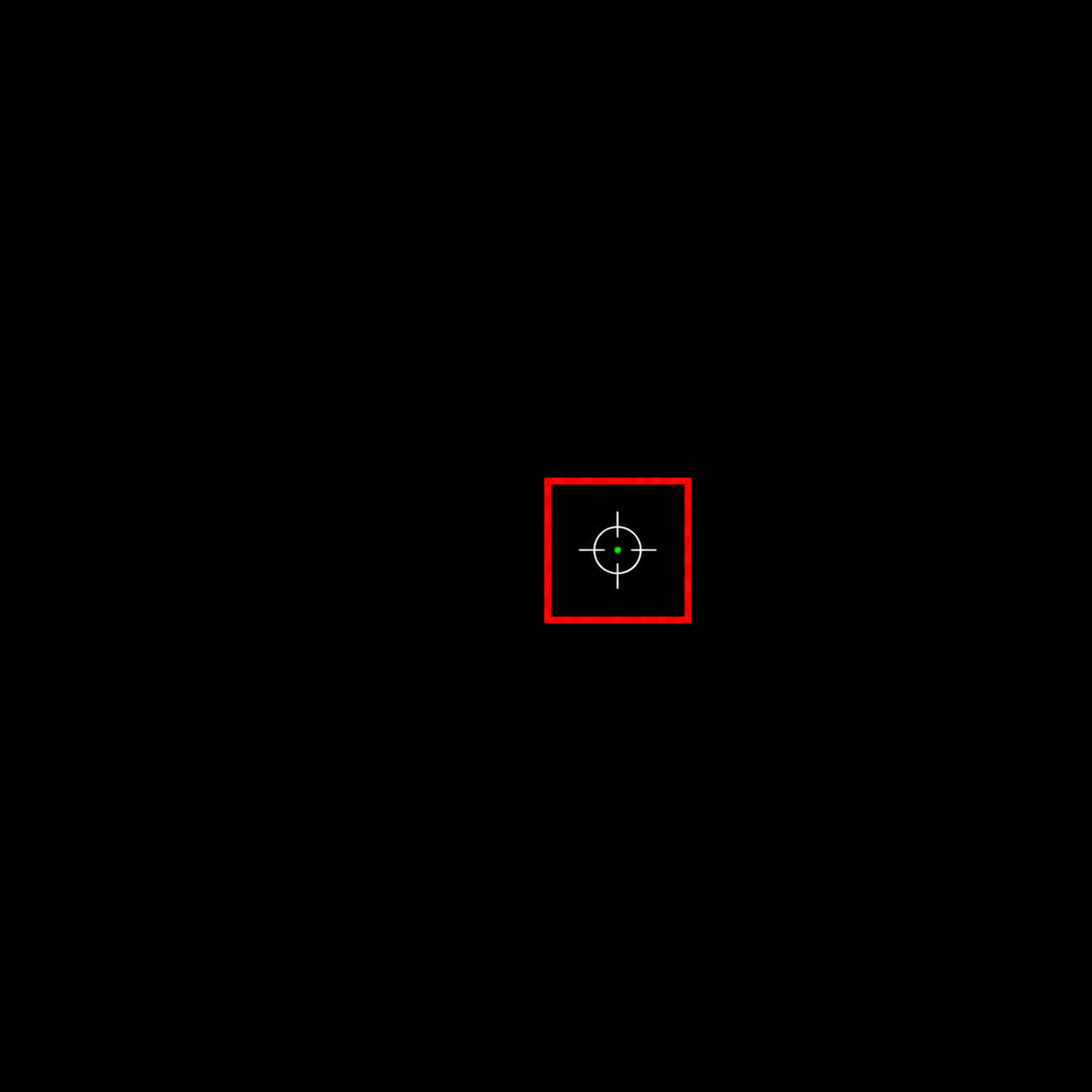}
        \caption{Antenna}
        \label{fig:modality_antenna}
    \end{subfigure}
    \hfill
    \begin{subfigure}[b]{0.48\columnwidth}
        \centering
        \includegraphics[width=\textwidth]{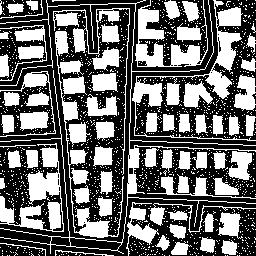}
        \caption{Mask}
        \label{fig:modality_mask}
    \end{subfigure}

    \vspace{4pt}

    \begin{subfigure}[b]{0.48\columnwidth}
        \centering
        \includegraphics[width=\textwidth]{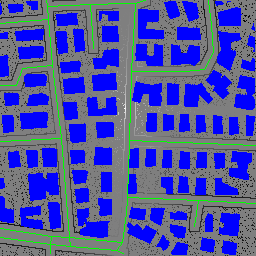}
        \caption{Measurement}
        \label{fig:modality_measurement}
    \end{subfigure}
    \hfill
    \begin{subfigure}[b]{0.48\columnwidth}
        \centering
        \includegraphics[width=\textwidth]{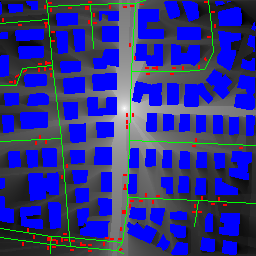}
        \caption{Color Radio Map}
        \label{fig:modality_radiomap}
    \end{subfigure}

    \caption{Input modalities and target output of the proposed RadioVIL framework. Subfigures show the antenna or structural layout, the physical mask defining observable and hidden regions, the degraded sparse measurement, and the complete color radio map used as the reconstruction target.}
    \label{fig:input_modalities}
\end{figure}

Given $\bm{y}$ and the pre-trained diffusion prior, our objective is to recover $\mathbf{R}^*$ and extract hidden vehicle coordinates. Fig.~\ref{fig:input_modalities} summarizes the task setting: the antenna and static structural layout specify the propagation environment, the mask defines the accessible sensing domain, the measurement provides sparse radio observations, and the color radio map is the dense target field to be reconstructed.

\begin{figure*}[!t]
    \centering
    \includegraphics[width=0.88\linewidth]{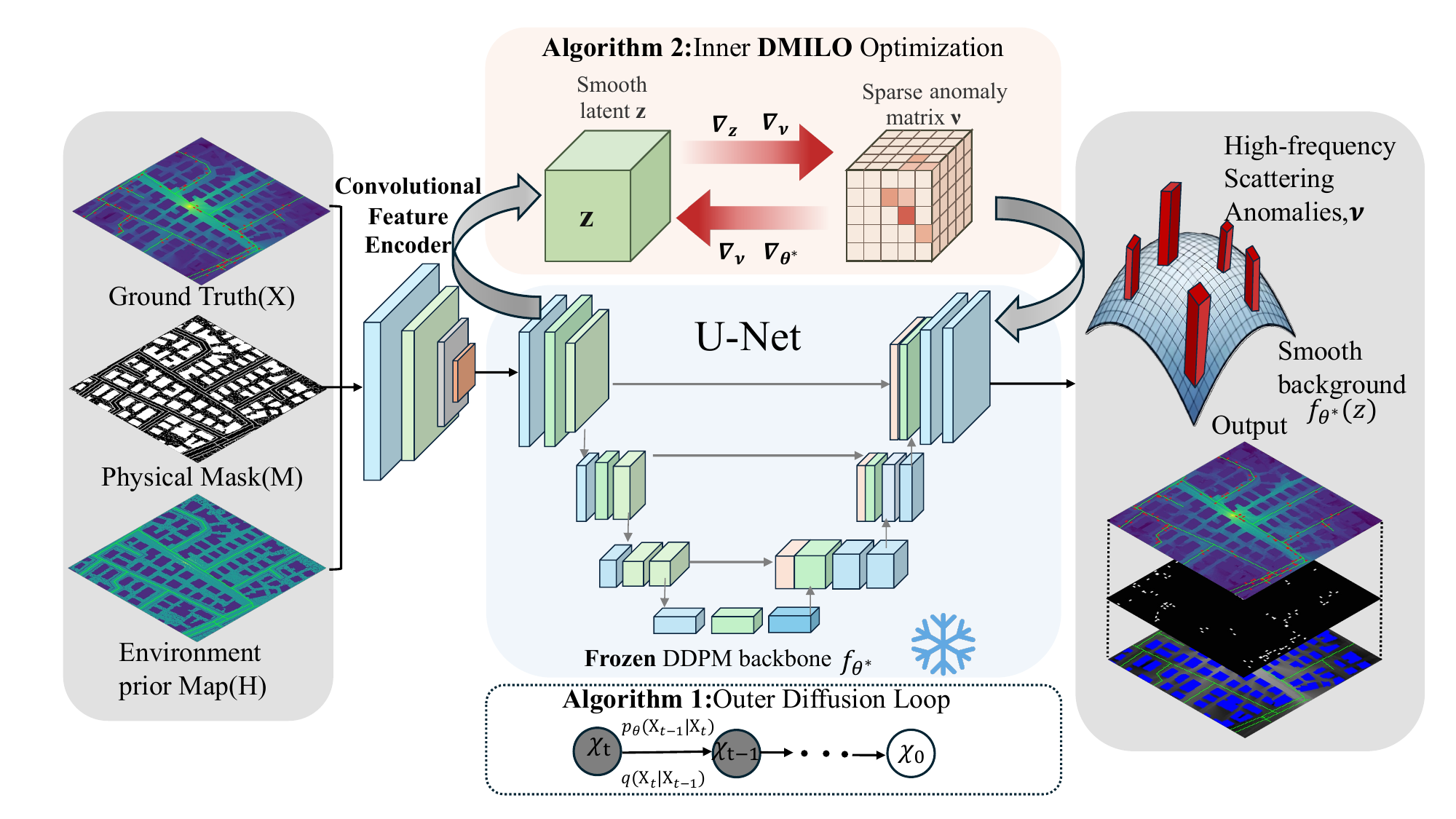}
    \caption{Overview of RadioVIL. The outer diffusion loop governs the sampling trajectory, while the inner DMILO loop minimizes $\mathcal{J}_t$. Through $L_1$-regularized optimization of $\bm{\nu}$, RadioVIL decouples the smooth background manifold $f_{\theta^*}(\bm{z})$ from sparse scattering anomalies for target localization.}
    \label{fig-system}
\end{figure*}

\section{Proposed RadioVIL Framework}
\label{sec:framework}

Fig.~\ref{fig-system} presents RadioVIL as a constrained bilevel pipeline combining a structural prior with anomaly-aware measurements.

\subsection{Prior Adaptation and Layer-wise Deterministic Mapping}

Offline, we train a DDPM backbone $f_{\theta^*}(\bm{x}_t, t, \mathbf{H})$, conditioned on the structural environment map $\mathbf{H}$, to model macroscopic radio propagation. By explicitly encoding building layouts and boundaries, this frozen prior effectively captures large-scale distance-dependent attenuation and shadowing. However, it lacks the capability to natively represent unpredictable, highly sparse scattering anomalies caused by dynamic vehicles. Guiding the reverse inference $\mathcal{G}(\cdot)$ directly with sparse measurements $\bm{y}$ requires unfolding the full computational graph, incurring prohibitive memory costs and disrupting the trajectory. To avoid this, we adopt a layer-wise DMILO strategy \cite{dmilo}. At each reverse timestep $t$, we introduce a trainable latent variable $\bm{z}$ to locally balance the generative prior and measurement constraints before proceeding.

\subsection{Anomaly-Aware Inversion Operator}

Sparse measurements $\bm{y}$ contain high-frequency residuals induced by vehicle scattering, which are difficult to represent using the smooth DDPM prior alone. To decouple these two factors, we introduce a sparse deviation variable $\bm{\nu}\in\mathbb{R}^{H\times W}$. At timestep $t$, the predicted measurement is
\begin{equation}
\hat{\bm{y}} =
\mathbf{M} \odot
\left(
f_{\theta^*}(\bm{z}, t, \mathbf{H})
+
\mathcal{K}(\bm{\nu})
\right),
\label{eq:predicted_measurement}
\end{equation}
where $f_{\theta^*}(\bm{z}, t, \mathbf{H})$ models the smooth background and $\mathcal{K}(\bm{\nu})$ describes the scattering footprint induced by sparse hidden anomalies. This leads to the layer-wise objective
\begin{equation}
\mathcal{J}_t(\bm{z}, \bm{\nu}) =
\frac{1}{2\sigma_n^2}
\left\|
\bm{y}
-
\mathbf{M} \odot
\left(
f_{\theta^*}(\bm{z}, t, \mathbf{H})
+
\mathcal{K}(\bm{\nu})
\right)
\right\|_F^2
+
\lambda \|\bm{\nu}\|_1.
\label{eq:dmilo_objective}
\end{equation}
Here, $\bm{z}$ adapts the smooth generative manifold to observed large-scale propagation, while $\bm{\nu}$ explains sparse residuals that remain inconsistent with this prior. Even when vehicle pixels are unobserved, $\bm{\nu}$ is inferable because $\mathcal{K}(\bm{\nu})$ affects the observed boundary samples. The $L_1$ penalty suppresses arbitrary residual fitting and encourages localized vehicle-induced anomalies. In this sense, $\bm{\nu}$ does not simply hallucinate missing pixels; instead, it serves as a sparse inverse variable constrained by both the measured boundary evidence and the smooth DDPM manifold.

\subsection{Iterative Inversion and Zero-Shot Extraction}

The procedure iteratively optimizes \eqref{eq:dmilo_objective}. At each $t$, we initialize $\bm{z} \leftarrow \bm{x}_t$ and $\bm{\nu} \leftarrow \mathbf{0}$, followed by $K$ gradient descent iterations to update the variables simultaneously:
\begin{align}
\bm{z} &\leftarrow \bm{z} - \eta_z \nabla_{\bm{z}} \mathcal{J}_t(\bm{z}, \bm{\nu}), \\
\bm{\nu} &\leftarrow \bm{\nu} - \eta_\nu \nabla_{\bm{\nu}} \mathcal{J}_t(\bm{z}, \bm{\nu}).
\end{align}
Upon convergence of the inner loop, the optimized state $\bm{z}^*$ is passed to the standard sampling step to compute $\bm{x}_{t-1}$, and the defect $\bm{\nu}^*$ is accumulated as $\hat{\bm{\nu}} = \hat{\bm{\nu}} + \bm{\nu}^*$.

After the diffusion loop concludes, we obtain the high-fidelity background radio map $\mathbf{R}_{est} = \bm{x}_0$ and the aggregated anomaly map $\hat{\bm{\nu}}$. A threshold $\tau$ generates a binary activation mask $\mathbf{B}_{veh}$:
\begin{equation}
\mathbf{B}_{veh}(p) =
\begin{cases}
1, & \text{if } |\hat{\bm{\nu}}(p)| > \tau \\
0, & \text{otherwise}
\end{cases}.
\label{eq:thresholding}
\end{equation}
By performing Connected Component Analysis (CCA) on $\mathbf{B}_{veh}$, we effectively filter out isolated noise pixels. Each robustly isolated cluster $\mathcal{C}_k$ corresponds to a physical vehicle. The continuous coordinates $(x_k, y_k)$ are derived by computing the center of mass based on the anomaly intensity:
\begin{equation}
x_k = \frac{\sum_{\mathcal{C}_k} x \cdot |\hat{\bm{\nu}}(x,y)|}{\sum_{\mathcal{C}_k} |\hat{\bm{\nu}}(x,y)|}, \quad y_k = \frac{\sum_{\mathcal{C}_k} y \cdot |\hat{\bm{\nu}}(x,y)|}{\sum_{\mathcal{C}_k} |\hat{\bm{\nu}}(x,y)|}.
\end{equation}
This mechanism enables highly accurate zero-shot vehicle localization directly from the physically isolated anomalies, eliminating the need for paired labels or auxiliary detection networks.

\section{Experimental Results and Analysis}
\label{sec:experiments}

\begin{table*}[!t]
\centering
\caption{Quantitative Comparison of Radio Map Inpainting and Zero-Shot Vehicle Localization Performance}
\label{tab:performance_comparison}

\vspace{2pt}
\footnotesize
\setlength{\tabcolsep}{4.2pt}
\renewcommand{\arraystretch}{1.22}

\begin{tabular}{@{}l ccc | cccc@{}}
\toprule
\multirow{2}{*}{\textbf{Method}}
& \multicolumn{3}{c|}{\textbf{Inpainting Quality}}
& \multicolumn{4}{c}{\textbf{Vehicle Localization Performance}} \\
\cmidrule(lr){2-4} \cmidrule(lr){5-8}
& \makecell{\textbf{PSNR (dB)}\\($\uparrow$)}
& \makecell{\textbf{SSIM}\\($\uparrow$)}
& \makecell{\textbf{LPIPS}\\($\downarrow$)}
& \makecell{\textbf{Recall (\%)}\\($\uparrow$)}
& \makecell{\textbf{F1-Score}\\($\uparrow$)}
& \makecell{\textbf{Mean Dist.}\\\textbf{(m)} ($\downarrow$)}
& \makecell{\textbf{P90 Dist.}\\\textbf{(m)} ($\downarrow$)} \\
\midrule
RadioUNet \cite{levie2021radiounet}
& 25.13 & 0.889 & 0.2107 & 0.00 & 0.00 & N/A & N/A \\

RME-GAN \cite{zhang2023rme}
& \textbf{25.50} & \textbf{0.9433} & 0.0889 & 0.00 & 0.00 & N/A & N/A \\

\rowcolor{gray!10}
RePaint \cite{Lugmayr2022RePaint}
& 23.35 & 0.8099 & 0.1010 & 11.44 & 0.1765 & 3.89 & \textbf{6.41} \\

\rowcolor{red!10}
\textbf{RadioVIL (Ours)}
& 24.30 & 0.8731 & \textbf{0.0587}
& \textbf{75.20} & \textbf{0.6799} & \textbf{3.31} & 7.40 \\

\rowcolor{blue!6}
\textbf{Relative Gain vs. RePaint}
& \textbf{+4.07\%}
& \textbf{+7.80\%}
& \textbf{41.88\%$\downarrow$}
& \textbf{+557.34\%}
& \textbf{+285.21\%}
& \textbf{14.91\%$\downarrow$}
& -- \\
\bottomrule
\end{tabular}

\vspace{2pt}
\begin{minipage}{0.96\textwidth}
\footnotesize
\textit{Note:} Relative gain is computed against RePaint; $\downarrow$ denotes relative reduction.
RePaint's P90 is computed on only 11.44\% successful detections, so its relative P90 gain is not reported.
\end{minipage}
\end{table*}

\begin{figure*}[!t]
\centering
\setlength{\tabcolsep}{0pt}
\renewcommand{\arraystretch}{0.95}
\begin{tabular}{ccccc}
\includegraphics[width=0.19\linewidth]{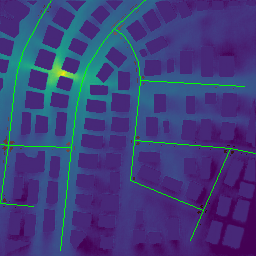} &
\includegraphics[width=0.19\linewidth]{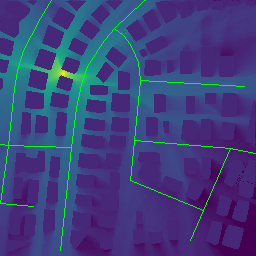} &
\includegraphics[width=0.19\linewidth]{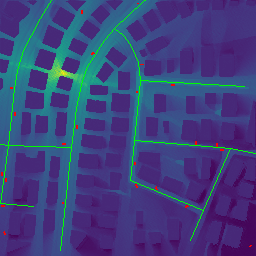} &
\includegraphics[width=0.19\linewidth]{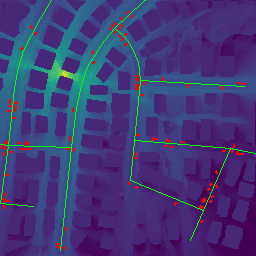} &
\includegraphics[width=0.19\linewidth]{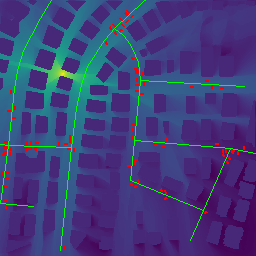} \\[-1pt]
\includegraphics[width=0.19\linewidth]{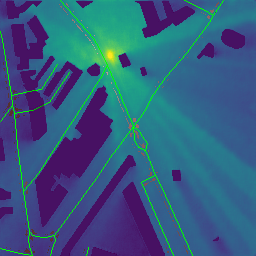} &
\includegraphics[width=0.19\linewidth]{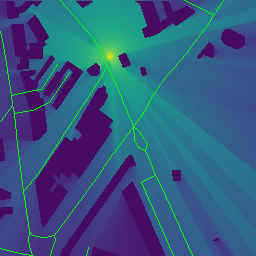} &
\includegraphics[width=0.19\linewidth]{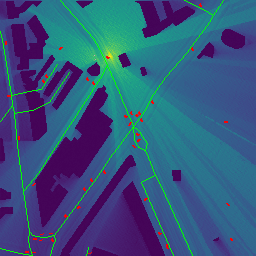} &
\includegraphics[width=0.19\linewidth]{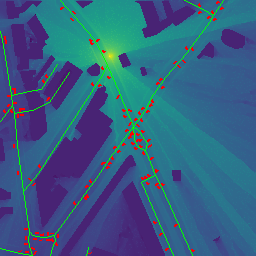} &
\includegraphics[width=0.19\linewidth]{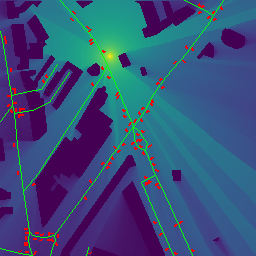} \\[-1pt]
\includegraphics[width=0.19\linewidth]{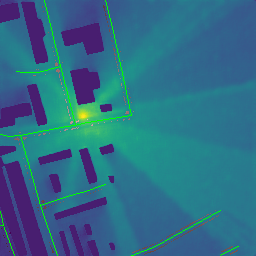} &
\includegraphics[width=0.19\linewidth]{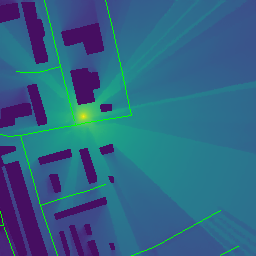} &
\includegraphics[width=0.19\linewidth]{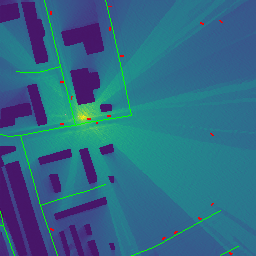} &
\includegraphics[width=0.19\linewidth]{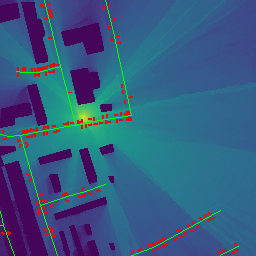} &
\includegraphics[width=0.19\linewidth]{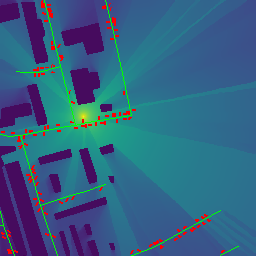} \\[-1pt]
\includegraphics[width=0.19\linewidth]{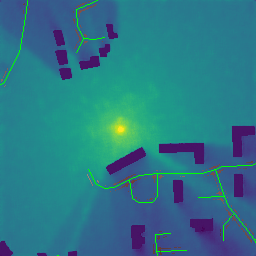} &
\includegraphics[width=0.19\linewidth]{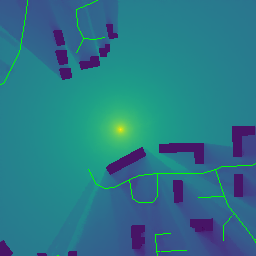} &
\includegraphics[width=0.19\linewidth]{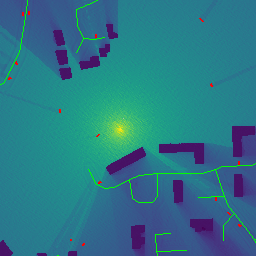} &
\includegraphics[width=0.19\linewidth]{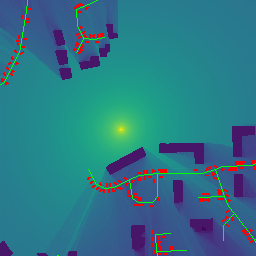} &
\includegraphics[width=0.19\linewidth]{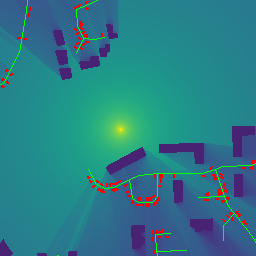} \\
RadioUNet &
RME-GAN &
RePaint &
RadioVIL (Ours) &
Ground Truth
\end{tabular}
\caption{Visual comparison of radio map inpainting and vehicle localization across different scenes from the RadioMapSeer dataset. From left to right: RadioUNet, RME-GAN, RePaint, RadioVIL, and ground truth.}
\label{fig-visual-results}
\end{figure*}

\subsection{Experimental Setup}

We evaluate RadioVIL on the RadioMapSeer benchmark dataset \cite{levie2021radiounet, yapar2023first}. To avoid target leakage, the conditioning map contains only static environmental information, including buildings, roads, greenbelts, and base-station-related cues; vehicle locations are excluded from the model input and used only for evaluation. To mimic sparse crowdsourced sensing, we retain static infrastructure, base-station coordinates, a single-pixel observation band around greenbelts, and 5\%--10\% randomly sampled free-space pixels. Vehicle regions are fully masked with $\mathbf{M}(p)=0$, forcing the method to infer hidden vehicles from boundary scattering residuals. This setting ensures that the localization task is not solved by direct visual access to vehicle regions, but by exploiting the indirect electromagnetic perturbations left in the observable measurements.

We compare RadioVIL against three representative deep learning paradigms:
\begin{itemize}
    \item \textbf{RadioUNet \cite{levie2021radiounet}:} A state-of-the-art fully supervised regression model optimized for pixel-wise Mean Squared Error (MSE).
    \item \textbf{RME-GAN \cite{zhang2023rme}:} A Generative Adversarial Network designed to hallucinate missing radio map data through adversarial training.
    \item \textbf{RePaint \cite{Lugmayr2022RePaint}:} A powerful zero-shot diffusion-based inpainting approach that conditions the unconditional generative process by forcibly replacing known pixels at each reverse timestep, aiming to harmoniously blend missing regions with the observed background.
\end{itemize}

The radio map inpainting quality is quantitatively assessed using Peak Signal-to-Noise Ratio (PSNR), Structural Similarity Index (SSIM), and Learned Perceptual Image Patch Similarity (LPIPS). For hidden vehicle localization, a predicted centroid is designated as a True Positive (TP) if its Euclidean distance to a ground-truth centroid falls within a specific tolerance limit (set to 10 meters in our evaluation). We employ Recall to measure the proportion of successfully identified true vehicles out of all actual targets, and the F1-Score as a comprehensive harmonic mean to evaluate precision and robustness. Furthermore, we assess spatial accuracy using the Mean Distance, calculating the average Euclidean coordinate error in meters, and introduce the 90th Percentile Distance (P90) to strictly bound the maximum spatial error for 90\% of the successful detections. It is important to note that these distance metrics are computed exclusively over the true positives.

During implementation, we employ the Adam optimizer for the layer-wise gradient descent. The learning rate for the mediating latent state $\bm{z}$ is set to $2\times 10^{-2}$, while the learning rate for the sparse deviation variable $\bm{\nu}$ is configured at $1\times 10^{-3}$. The number of inner iterations is defined as $K = 200$, and the outer diffusion loop operates for 5 reverse timesteps. To forcefully enforce physical constraints, the sparsity regularization weight is strictly fixed at $\lambda = 0.1$, and the anomaly activation threshold for extracting zero-shot vehicle masks is empirically established at $\tau = 0.5$.

\subsection{Radio Map Inpainting Performance}

The quantitative results reveal an important phenomenon in metric evaluation. Although traditional regression and GAN-based baselines (RadioUNet, RME-GAN) achieve higher PSNRs (25.13 dB and 25.50 dB), PSNR tends to penalize high-frequency deviations, encouraging overly smoothed reconstructions that erase vehicle scattering footprints.

Compared with the zero-shot diffusion baseline RePaint, RadioVIL shows clear advantages, achieving higher PSNR (24.30 dB vs. 23.35 dB), SSIM, and the best LPIPS score of 0.0587. Since RePaint is designed for visual semantic harmonization, it treats sharp physical scattering boundaries as visual noise and blends critical target signatures into the background. In contrast, RadioVIL separates the smooth environmental prior from scattering deviations through layer-wise DMILO optimization, preserving physical textures essential for sensing.

\subsection{Vehicle Localization Performance}

The handling of high-frequency anomalies directly determines vehicle localization performance. As shown in Table~\ref{tab:performance_comparison}, traditional methods fail catastrophically: both RadioUNet \cite{levie2021radiounet} and RME-GAN \cite{zhang2023rme} detect no hidden vehicles, yielding F1-scores of 0. RePaint \cite{Lugmayr2022RePaint} performs only marginally better, with 11.44\% Recall and an F1-Score of 0.1765.

Although RePaint reports a lower 90th Percentile Distance (6.41m vs. 7.40m), this value is computed over only 11.44\% successful detections. Therefore, P90 should be interpreted jointly with Recall and F1-Score, where RadioVIL bounds the localization error for a much larger portion (75.20\%) of hidden vehicles. As shown in Fig.~\ref{fig-visual-results}, RePaint suppresses structural anomalies and blends masked regions into free space. Since it is not anomaly-aware, it treats high-frequency electromagnetic scattering footprints as noise, causing frequent missed detections.

In contrast, RadioVIL achieves robust localization under sparse and strictly masked observations. By applying zero-shot extraction to the isolated anomaly map $\hat{\bm{\nu}}$, our framework obtains an F1-Score of 0.6799, recalls 75.20\% of hidden vehicles, and achieves an average localization error of 3.31 meters. These results demonstrate that RadioVIL bridges visual image completion and anomaly-aware physical environment sensing.

\subsection{Robustness to Environmental Factors}

\begin{table}[!t]
\centering
\caption{Impact of Building Density on Performance}
\label{tab:building_impact}
\setlength{\tabcolsep}{3.5pt}
\renewcommand{\arraystretch}{1.2}
\begin{tabular}{l c c c c}
\toprule
\makecell{\textbf{Buildings}} & \textbf{Samples} & \makecell{\textbf{PSNR}} & \textbf{SSIM} & \makecell{\textbf{F1}} \\
\midrule
$< 15$ (Simple) & 560 & 24.55 & 0.878 & 0.743 \\
$15 - 30$ & 5440 & 24.12 & 0.868 & 0.675 \\
$> 30$ (Complex) & 2000 & 24.56 & 0.874 & 0.681 \\
\bottomrule
\end{tabular}
\end{table}

We investigate the robustness of RadioVIL against environmental complexity. As summarized in Table~\ref{tab:building_impact}, performance shows no clear monotonic degradation as the environment becomes more cluttered, indicating that RadioVIL does not rely on overly simplified propagation patterns or sparse building layouts. Even in dense urban scenes with more than 30 buildings, where blockage, shadowing, and multipath effects are more severe, the vehicle detection F1-Score decreases by only 8.3\%. This suggests that RadioVIL can still exploit global structural priors and boundary scattering residuals under complex propagation conditions.

\begin{figure}[!t]
    \centering
    \includegraphics[width=0.8\columnwidth]{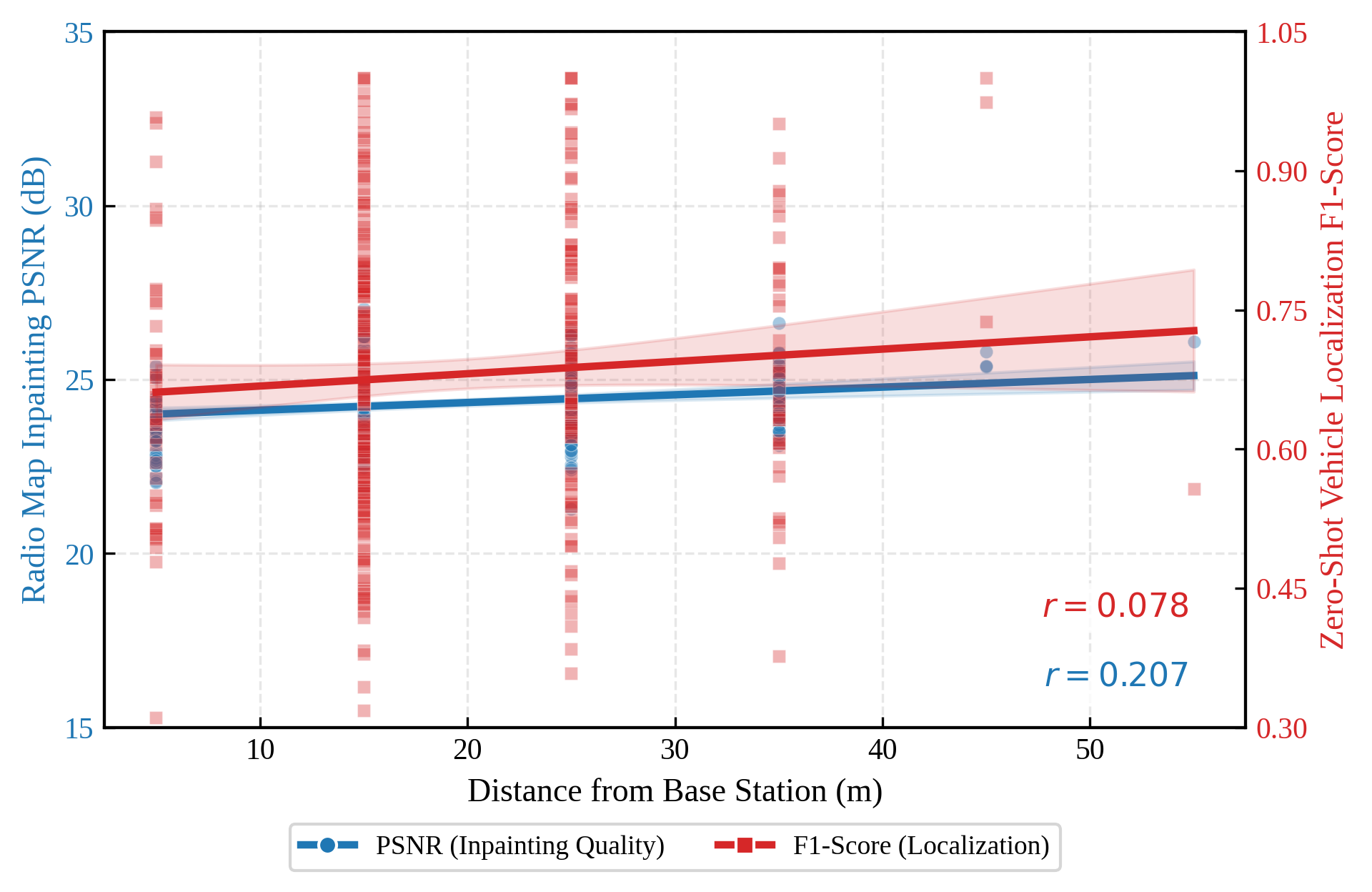}
    \caption{Performance stability vs. distance from base station.}
    \label{fig:distance_trend}
\end{figure}

Furthermore, Fig.~\ref{fig:distance_trend} shows that both radio map inpainting and vehicle localization remain relatively stable across transmission ranges from 50m to 800m. This indicates that RadioVIL relies on global environmental structural priors rather than local signal intensity alone, enabling reliable performance across the coverage area.

\section{Conclusion}

This paper presented RadioVIL, an anomaly-aware diffusion framework for joint radio map inpainting and zero-shot hidden vehicle localization from sparse measurements. By combining a frozen DDPM prior with layer-wise DMILO optimization, RadioVIL reformulates radio map reconstruction as a prior-guided physical inverse problem rather than a conventional pixel-level completion task. The frozen diffusion model captures the smooth environmental propagation background, while the $L_1$-regularized sparse deviation variable isolates vehicle-induced scattering anomalies that cannot be explained by the smooth prior. This design enables RadioVIL to preserve high-frequency electromagnetic signatures that are often suppressed by regression-based, GAN-based, or semantic diffusion inpainting baselines.

The isolated anomaly map further enables CCA-based vehicle localization without paired vehicle labels or auxiliary detection networks, supporting zero-shot sensing under strict masking conditions. Overall, the results demonstrate that sparse wireless measurements, when guided by structural generative priors and anomaly-aware inverse optimization, can provide a promising sensing mechanism for future 6G ISAC systems.

\section*{Acknowledgment}
This work was supported by the National Key Research and Development Program of China (2024YFB907500), and the National Natural Science Foundation of China (62571402).

\balance

\bibliographystyle{IEEEtran}
\bibliography{ref}

\end{document}